\documentclass[12pt]{iopart}

\usepackage{amssymb}
\usepackage{bm}
\usepackage{bbold}

\begin{document}

\title[Incomplete Dirac reduction]{Incomplete Dirac reduction of constrained Hamiltonian systems}

\author{C. Chandre}
\address{Centre de Physique Th\'eorique, CNRS -- Aix-Marseille Universit\'e, Campus de Luminy, case 907, 13009 Marseille, France}
\ead{chandre@cpt.univ-mrs.fr}

\begin{abstract}
First-class constraints constitute a potential obstacle to the computation of a Poisson bracket in Dirac's theory of constrained Hamiltonian systems. Using the pseudoinverse instead of the inverse of the matrix defined by the Poisson brackets between the constraints, we show that a Dirac-Poisson bracket can be constructed, even if it corresponds to an incomplete reduction of the original Hamiltonian system. The uniqueness of Dirac brackets is discussed. 
\end{abstract}

\maketitle



\section{Introduction}

We consider an $N$-dimensional Hamiltonian system with phase space variables ${\bf z}=(z_1,z_2,\ldots,z_N)$ given by its Hamiltonian $H({\bf z})$ and its Poisson bracket $\{\cdot,\cdot\}$. We impose a set of $K$ constraints $\Phi_n({\bf z})=0$ for $n=1,\ldots,K$ on this dynamical system. We define the Dirac bracket as
\begin{equation}
\label{eqn:gDB}
\{F,G\}_*=\{F,G\}-\{F,\Phi_n\}D_{nm}\{\Phi_m,G\},
\end{equation}
with implicit summation over repeated indices.
Usually the matrix ${\mathbb D}$ whose elements are $D_{nm}$ is taken as the inverse of the matrix ${\mathbb C}$ whose elements are 
$$
C_{nm}=\{\Phi_n,\Phi_m\},
$$
if it is invertible~\cite{dira50,dira58,Bhans76,Bsund82,Bhenn92,wipf94}. In this case, it has been shown that the usual Dirac bracket~(\ref{eqn:gDB}) is a Poisson bracket~\cite{dira50}, in particular, that it satisfies the Jacobi identity everywhere in phase space (and not just on the surface defined by the constraints). Dirac's theory of constrained Hamiltonian systems has been used in a wide variety of contexts~\cite{wipf94}. Recently, it has been used to derived reduced models in fluid and plasma physics~\cite{morr09,chan10,chan12,chan13}. 

In the literature there are some tentative definitions of weak Dirac bracket~\cite{deri96,bizd00,abre02,bizd07} with the aim of defining a Dirac bracket even in situations where the matrix ${\mathbb C}$ is not invertible. The non-invertibility of ${\mathbb C}$ is linked to the existence of first class constraints, i.e., constraints that commute with all the other constraints (in the weak sense). 
The tentatives to define such a generalization of the Dirac bracket so far did not result in the definition of a well defined Poisson bracket, i.e., which satisfies the Jacobi identity everywhere in phase space. More specifically, in these tentatives, the Jacobi identity together with the commutation of the constraints with any function only happens on the surface defined by the constraints. 

Here we generalize the Dirac bracket to cases where ${\mathbb C}$ is not invertible by taking ${\mathbb D}$ as the Moore-Penrose pseudoinverse of ${\mathbb C}$.  The conditions which ${\mathbb D}$ has to satisfy are
\begin{eqnarray}
&& {\mathbb C}={\mathbb C}{\mathbb D}{\mathbb C},\label{cond1}\\
&& {\mathbb D}={\mathbb D}{\mathbb C}{\mathbb D},\label{cond2}\\
&& {\mathbb C}{\mathbb D} = {\mathbb D}{\mathbb C}. \label{cond3}
\end{eqnarray}
In finite dimensions, the pseudoinverse always exists, so a Dirac bracket of the form~(\ref{eqn:gDB}) can always be computed regardless of the constraints and the original Poisson bracket. 
If the Poisson bracket is given by
\begin{equation}
\label{eq:FGJ}
\{F,G\}= \frac{\partial F}{\partial {\bf z}}\cdot {\mathbb J}({\bf z})\frac{\partial G}{\partial{\bf z}},
\end{equation}
where ${\mathbb J}$ is the Poisson matrix, the Dirac bracket $\{\cdot,\cdot\}_*$ has the same expression as Eq.~(\ref{eq:FGJ}) where ${\mathbb J}$ is replaced by
\begin{equation}
{\mathbb J}_*={\mathbb J}-{\mathbb J}\hat{\cal Q}^\dagger {\mathbb D}\hat{\cal Q} {\mathbb J}, \label{eqn:Jstarmat}
\end{equation}
where $\hat{\cal Q}$ has elements $\hat{\cal Q}_{ni}=\partial \Phi_n/\partial z_i$.
The linear operator ${\cal P}_*=1-\hat{\cal Q}^\dagger {\mathbb D}\hat{\cal Q} {\mathbb J}$ is a projector, called Dirac projector~\cite{chan13}. In order to prove this we need condition~(\ref{cond2}). From this projector, we have the following identities~:
$$
{\mathbb J}_*={\cal P}_*^\dagger {\mathbb J}{\cal P}_*={\mathbb J}{\cal P}_*,
$$
which means that the Dirac bracket is the same as the original bracket with the exception that the derivatives $\partial F/\partial {\bf z}$ in the Poisson bracket have to be replaced by the constrained derivatives defined by
$$
\left. \frac{\partial F}{\partial {\bf z}}\right|_{\rm c} ={\cal P}_* \frac{\partial F}{\partial {\bf z}}. 
$$ 
The constraints $\Phi_n({\bf z})$ are Casimir invariants of the Dirac bracket, i.e., $\{\Phi_n({\bf z}),G\}_*=0$ for all functionals $G$, if and only if ${\mathbb J}_*\hat{\cal Q}^\dagger=0$. Therefore a complete reduction corresponds to the case where all constraints are Casimir invariants of the bracket~(\ref{eqn:gDB}). However even if the Dirac bracket can always be computed (at least for finite dimensional systems), ${\mathbb J}_*\hat{\cal Q}={\mathbb J}\hat{\cal Q}(1-{\mathbb D}{\mathbb C})$ is non-zero in general. We will see below that the obstacle to the complete reduction corresponds to the primary constraints which are not Casimir invariants of the original bracket $\{\cdot,\cdot\}$.

It has been shown in Ref.~\cite{chan13} that if the reduction is complete, then the Dirac bracket~(\ref{eqn:gDB}) is a Poisson bracket, i.e., it satisfies the Jacobi identity everywhere in phase space. Here we prove that, even in the case of an incomplete reduction, the generalized Dirac bracket defined by Eq.~(\ref{eqn:gDB}) with the conditions~(\ref{cond1})-(\ref{cond3}) is still a Poisson bracket, i.e., in Sec.~\ref{sec:proof} we prove that the bracket~(\ref{eqn:gDB}) satisfies the Jacobi identity everywhere in phase space. In Sec.~\ref{sec:example}, we apply the generalized Dirac bracket to several examples, finite and infinite dimensional ones.

\section{Proof of the Jacobi identity}
\label{sec:proof}

First, we recall that first-class constraints are constraints which Poisson-commute with all the other constraints for the original bracket  $\{\cdot ,\cdot \}$. In particular, constraints which are Casimir invariants of the original bracket are first-class constraints, but obviously first-class constraints are not restricted to Casimir invariants of the original Poisson bracket. 

Using a local change of variables, we use the constraints as part of the variables, i.e., we assume $\Phi_k({\bf z})=z_k$ for $k=1,\ldots, K$. We divide this set of constraints/variables into three groups: 
\begin{itemize}
\item the first-class constraints which are not Casimir invariants (of the original bracket), for $k=1,\ldots, k_1$,
\item the (first-class) constraints which are Casimir invariants, for $k=k_1+1,\ldots, k_1+k_c$,
\item the second-class constraints, for $k=k_1+k_c+1,\ldots, k_1+k_c+k_2$,
\end{itemize}
where $K=k_1+k_c+k_2$. The arrangement of the constraints is done such that there is no linear combination of second-class constraints which is a first-class constraint. The rest of the variables are unchanged. 
Using this partition of the variables, we rewrite the Poisson matrix ${\mathbb J}$ as
$$
{\mathbb J}=\left( \begin{array}{cccc} 0 & 0 & 0 & -c^\dagger \\
0 & 0 & 0 & 0\\ 0 & 0 & {\mathbb A} & -{\mathbb B}^\dagger \\
c & 0 & {\mathbb B} & \bar{\mathbb J}\end{array}\right).
$$
Since $z_k$ for $k=1,\ldots, k_1$ are first-class constraints, the first column has three zeros, but the last element is non-zero (otherwise they would be Casimir invariants). The second column is zero since $z_k$ are Casimir invariants for $k=k_1+1,\ldots, k_1+k_c$.    
The matrix ${\mathbb A}$ is invertible, since otherwise there would exist a linear combination of the second-class constraints which would be a first-class constraints, which is excluded by construction. 
The operator $\hat{\cal Q}$ is given by
$$
\hat{\cal Q}=\left( \begin{array}{cccc} {\mathbb 1}_{k_1} & 0 & 0 & 0\\
0 & {\mathbb 1}_{k_c} & 0 & 0\\ 0 & 0 & {\mathbb 1}_{k_2} & 0
\end{array}\right),
$$
where ${\mathbb 1}_k$ is the $k$-dimensional identity matrix. 
The matrix ${\mathbb C}=\hat{\cal Q}{\mathbb J}\hat{\cal Q}^\dagger$ is given by
$$
{\mathbb C}=\left( \begin{array}{ccc} 0 & 0 & 0\\
0 & 0 & 0\\
0 & 0 & {\mathbb A}\end{array}\right).
$$
The pseudoinverse of ${\mathbb C}$ is given by 
$$
{\mathbb D}=\left( \begin{array}{ccc} 0 & 0 & 0\\
0 & 0 & 0\\
0 & 0 & {\mathbb A}^{-1}\end{array}\right).
$$
The Poisson matrix associated with the Dirac bracket~(\ref{eqn:gDB}) is obtained from Eq.~(\ref{eqn:Jstarmat}) as
\begin{equation}
\label{eqn:Jstar}
{\mathbb J}^*=\left( \begin{array}{cccc} 0 & 0 & 0 & -c^\dagger\\
0 & 0 & 0 & 0 \\ 0 & 0 & 0 & 0\\ c & 0 & 0 & \bar{\mathbb J}+{\mathbb B}{\mathbb A}^{-1}{\mathbb B}^\dagger \end{array}\right).
\end{equation}

From this expression, we readily see that the first-class constraints that are not Casimir invariants of the original bracket, are not Casimir invariants of the Dirac bracket. The only way to resolve these first-class constraints in a Hamiltonian way, i.e., to have these constraints as conserved quantities, is to modify the Hamiltonian (by removing the dependence of the Hamiltonian on the variables corresponding to first-class non-Casimir constraints). 

In this section, the goal is to prove the Jacobi identity for ${\mathbb J}_*$ given by Eq.~(\ref{eqn:Jstar}) under the hypothesis that ${\mathbb J}$ satisfies the Jacobi identity, i.e.,
\begin{equation}
\label{eqn:JacJ}
J_{il}\partial_l J_{jk}+\circlearrowleft_{(i,j,k)}=0,
\end{equation}
for all $(i,j,k)$. 
In what follows we denote $I_1$ the set of indices $k$ for which $z_k$ is a non-Casimir first-class constraints, $I_{\rm c}$ the set of indices such that $z_k$ is a Casimir invariant of ${\mathbb J}$, $I_2$ the set of indices such that $z_k$ is a second-class constraint, and $I_{\rm n}$ all the other indices. 
From Eq.~(\ref{eqn:JacJ}) there are seven non-trivial identities listed below:
\begin{itemize}

\item If $(i,j)\in I_1$ and $k\in I_{\rm n}$, 
\begin{equation}
\label{eqn:J11n}
c_{li}\partial_l c_{kj}-c_{lj}\partial_l c_{ki}=0.
\end{equation}

\item If $i\in I_1$ and $j,k \in I_2$,
\begin{equation}
\label{eqn:J122}
c_{li}\partial_l A_{jk}=0.
\end{equation}

\item If $i\in I_1$, $j\in I_2$ and $k\in I_{\rm n}$
\begin{equation}
\label{eqn:J12n}
c_{li}\partial_l B_{kj}+ A_{jl}\partial_l c_{ki}-B_{lj}\partial_l c_{ki}=0.
\end{equation}

\item If $i\in I_1$ and $j,k \in I_{\rm n}$
\begin{eqnarray}
&&-c_{li}\partial_l \bar{J}_{jk}-c_{kl}\partial_l c_{ji}+c_{jl}\partial_l c_{ki}
-B_{kl}\partial_l c_{ji}+B_{jl} \partial_l c_{ki}\nonumber \\
&& \qquad \qquad  -\bar{J}_{kl}\partial_l c_{ji}
+\bar{J}_{jl}\partial_l c_{ki}=0.\label{eqn:J1nn}
\end{eqnarray}

\item If $i,j,k\in I_{\rm n}$
\begin{equation}
\label{eqn:Jnnn}
c_{il}\partial_l \bar{J}_{jk}+B_{il}\partial_l \bar{J}_{jk}+\bar{J}_{il}\partial_l \bar{J}_{jk}+\circlearrowleft_{(i,j,k)}=0.
\end{equation}

\item If $i,j\in I_{\rm n}$ and $k\in I_2$
\begin{eqnarray}
&&c_{il}\partial_l B_{jk} -c_{jl}\partial_l B_{ik}+B_{il}\partial_l B_{jk}-B_{jl}\partial_l B_{ik} +\bar{J}_{il}\partial_l B_{jk}-\bar{J}_{jl}\partial_l B_{ik}\nonumber \\
&& \qquad \qquad  -B_{lk}\partial_l \bar{J}_{ij}+A_{kl}\partial_l \bar{J}_{ij}=0.\label{eqn:Jnn2}
\end{eqnarray}

\item If $i\in I_{\rm n}$ and $j,k \in I_2$
\begin{eqnarray}
&&c_{il}\partial_l A_{jk}+A_{kl}\partial_l B_{ij}-A_{jl}\partial_l B_{ik}+B_{il}\partial_l A_{jk} -B_{lk}\partial_l B_{ij}\nonumber \\
&& \qquad \qquad  +B_{lj}\partial_l B_{ik}+\bar{J}_{il}\partial_l A_{jk}=0.\label{eqn:Jn22}
\end{eqnarray}

\item If $i,j,k \in I_2$ 
\begin{equation}
\label{eqn:J222}
A_{il}\partial_l A_{jk}-B_{li}\partial_l A_{jk}+\circlearrowleft_{(i,j,k)}=0.
\end{equation}
\end{itemize}

We denote 
\begin{equation}
S_{ijk}=J^*_{il}\partial_l J^*_{jk}+\circlearrowleft_{(i,j,k)}.
\label{Jacobiator}
\end{equation}
Using Eqs.~(\ref{eqn:J11n})-(\ref{eqn:J222}), we prove below that $S_{ijk}=0$ for all $i,j,k$.
Several cases have to be envisaged, depending on which set $i$, $j$ and $k$ belong to. There are only two cases where the derivation of the Jacobi identity is non-trivial, one corresponding to $i,j,k\in I_{\rm n}$ and the other one corresponding to $i\in I_1$ and $j,k\in I_{\rm n}$. 

\begin{itemize}

\item If one of the indices $i,j,k$ belong to $I_2$ or $I_{\rm c}$, then all the terms $J^*_{il}\partial_l J^*_{jk}$ vanish individually and the Jacobi identity is trivially satisfied. Consequently $i,j,k$ belong to either $I_1$ or $I_{\rm n}$.

\item If $i,j,k\in I_1$, then $J_{jk}^*=0$ and hence the Jacobi identity is trivially satisfied. 

\item If $i,j\in I_1$ and $k\in I_{\rm n}$, 
$$
S_{ijk}=c_{li}\partial_l c_{kj}-c_{lj}\partial_l c_{ki},
$$ 
which vanishes considering Eq.~(\ref{eqn:J11n}).

\item If $i\in I_1$ and $j,k\in I_{\rm n}$, 
\begin{eqnarray}
S_{ijk}&=&-c_{li}\partial_l (\bar{\mathbb J}
+{\mathbb B}{\mathbb A}^{-1}{\mathbb B}^\dagger)_{jk} -c_{kl}\partial_l c_{ji}+c_{jl}\partial_l c_{ki}\nonumber \\
&& - (\bar{\mathbb J}
+{\mathbb B}{\mathbb A}^{-1}{\mathbb B}^\dagger)_{kl}\partial_l c_{ji} +(\bar{\mathbb J}
+{\mathbb B}{\mathbb A}^{-1}{\mathbb B}^\dagger)_{jl}\partial_l c_{ki}.\label{eqn:intermDB}
\end{eqnarray}
In order to show that $S_{ijk}=0$, we need Eqs.~(\ref{eqn:J122}), (\ref{eqn:J12n}) and (\ref{eqn:J1nn}). Using Eq.~(\ref{eqn:J1nn}), $S_{ijk}$ is rewritten as
$$
S_{ijk}=-c_{li}\partial_l ({\mathbb B}{\mathbb A}^{-1}{\mathbb B}^\dagger)_{jk}-
({\mathbb B}{\mathbb A}^{-1}{\mathbb B}^\dagger)_{kl}\partial_l c_{ji}+
({\mathbb B}{\mathbb A}^{-1}{\mathbb B}^\dagger)_{jl}\partial_l c_{ki}
+B_{kl}\partial_l c_{ji}-B_{jl}\partial_l c_{ki}.
$$
From ${\mathbb A}{\mathbb A}^{-1}={\mathbb 1}_{k_2}$, we have
$$
A_{\beta\gamma}^{-1}\partial_l A_{\alpha \beta}+A_{\alpha\beta}\partial_l A_{\beta\gamma}^{-1}=0.
$$
Multiplying the previous identity by $c_{li}$ and using Eq.~(\ref{eqn:J122}), we have
$$
c_{li}A_{\alpha\beta}\partial_l A_{\beta\gamma}^{-1}=0.
$$
Since ${\mathbb A}$ is invertible, this identity becomes equivalent to 
$$
c_{li}\partial_l A_{\beta\gamma}^{-1}=0,
$$
which is analogous to Eq.~(\ref{eqn:J122}) for ${\mathbb A}^{-1}$. In order to prove that $S_{ijk}=0$, we compute $c_{li}\partial_l ({\mathbb B}{\mathbb A}^{-1}{\mathbb B}^\dagger)_{jk}$:
$$
c_{li}\partial_l ({\mathbb B}{\mathbb A}^{-1}{\mathbb B}^\dagger)_{jk}=c_{li}\partial_l
B_{j\alpha}A_{\alpha\beta}^{-1}B_{k\beta}+c_{li}\partial_l B_{k\beta}A_{\alpha\beta}^{-1}
B_{j\alpha}+c_{li}B_{j\alpha}B_{k\beta}\partial_l A_{\alpha\beta}^{-1}.
$$
Using $c_{li}\partial_l A_{\alpha\beta}^{-1}=0$ and Eq.~(\ref{eqn:J12n}) to rewrite $c_{li}\partial_l B_{j\alpha}$ and $c_{li}\partial_l B_{k\beta}$, we obtain
$$
c_{li}\partial_l ({\mathbb B}{\mathbb A}^{-1}{\mathbb B}^\dagger)_{jk}=(-A_{\alpha l}\partial_l c_{ji}+B_{l\alpha}\partial_l c_{ji})A_{\alpha\beta}^{-1}B_{k\beta}
+(-A_{\beta l}\partial_l c_{ki}+B_{l\beta}\partial_l c_{ki})A_{\alpha\beta}^{-1}B_{j\alpha}.
$$
Since ${\mathbb A}$ is antisymmetric, we rewrite this expression as
$$
c_{li}\partial_l ({\mathbb B}{\mathbb A}^{-1}{\mathbb B}^\dagger)_{jk}=B_{kl}\partial_l c_{ji}-B_{jl}\partial_l c_{ki}
+({\mathbb B}{\mathbb A}^{-1}{\mathbb B}^\dagger)_{lk}\partial_l c_{ji}-({\mathbb B}{\mathbb A}^{-1}{\mathbb B}^\dagger)_{lj}\partial_l c_{ki},
$$
from which we deduce that $S_{ijk}=0$. 

\item The case $i,j,k\in I_{\rm n}$ is the most involved one. The expression of $S_{ijk}$ is given by
$$
S_{ijk}=c_{il}\partial_l (\bar{\mathbb J}
+{\mathbb B}{\mathbb A}^{-1}{\mathbb B}^\dagger)_{jk}+(\bar{\mathbb J}
+{\mathbb B}{\mathbb A}^{-1}{\mathbb B}^\dagger)_{il}\partial_l (\bar{\mathbb J}
+{\mathbb B}{\mathbb A}^{-1}{\mathbb B}^\dagger)_{jk}+\circlearrowleft_{(i,j,k)}.
$$
We start by inserting Eq.~(\ref{eqn:Jnnn}) into $S_{ijk}$~:
$$
S_{ijk}=c_{il}\partial_l ({\mathbb B}{\mathbb A}^{-1}{\mathbb B}^\dagger)_{jk}+
\bar{J}_{il}\partial_l ({\mathbb B}{\mathbb A}^{-1}{\mathbb B}^\dagger)_{jk}+
({\mathbb B}{\mathbb A}^{-1}{\mathbb B}^\dagger)_{il}\partial_l (\bar{\mathbb J}+{\mathbb B}{\mathbb A}^{-1}{\mathbb B}^\dagger)_{jk}-B_{il}\partial_l \bar{J}_{jk}+\circlearrowleft_{(i,j,k)}.
$$
Expanding the above expression leads to 
\begin{eqnarray*}
S_{ijk}&=&c_{il}\partial_l B_{j\alpha}A_{\alpha\beta}^{-1}B_{k\beta}+c_{il}B_{j\alpha}
A_{\alpha\beta}^{-1}\partial_l B_{k\beta}\\
&& +\bar{J}_{il}\partial_l B_{j\alpha}A_{\alpha\beta}^{-1}B_{k\beta}+\bar{J}_{il}B_{j\alpha}
A_{\alpha\beta}^{-1}\partial_l B_{k\beta}\\
&&+ B_{i\alpha}A_{\alpha\beta}^{-1}B_{l\beta}\partial_l B_{j\gamma}A_{\gamma\delta}^{-1} B_{k\delta}+ B_{i\alpha}A_{\alpha\beta}^{-1}B_{l\beta}B_{j\gamma}A_{\gamma\delta}^{-1}\partial_l B_{k\delta}\\
&& +c_{il}B_{j\alpha}B_{k\beta}\partial_l A_{\alpha\beta}^{-1}+\bar{J}_{il}B_{j\alpha}B_{k\beta}\partial_l A_{\alpha\beta}^{-1} +
B_{i\alpha}A_{\alpha\beta}^{-1}B_{l\beta}B_{j\gamma}B_{k\delta}\partial_l A_{\gamma\delta}^{-1}\\
&&+ B_{i\alpha}A_{\alpha\beta}^{-1}B_{l\beta}\partial_l \bar{J}_{jk}-B_{il}\partial_l \bar{J}_{jk}+\circlearrowleft_{(i,j,k)}.
\end{eqnarray*}
Using a circular permutation and antisymmetry of ${\mathbb A}^{-1}$ with which $c_{il}B_{j\alpha}A_{\alpha\beta}^{-1}\partial_l B_{k\beta}$ is replaced by $-c_{jl}\partial_l B_{i\alpha}A_{\alpha\beta}^{-1}B_{k\beta}$, the expression for $S_{ijk}$ becomes
\begin{eqnarray*}
S_{ijk}&=& (c_{il}\partial_l B_{j\alpha}-c_{jl}\partial_l B_{i\alpha}+\bar{J}_{il}\partial_l B_{j\alpha}-\bar{J}_{jl}\partial_l B_{i\alpha})A_{\alpha\beta}^{-1}B_{k\beta}\\
&& +B_{i\alpha}A_{\alpha\beta}^{-1}B_{l\beta}\partial_l B_{j\gamma}A_{\gamma\delta}^{-1} B_{k\delta}+ B_{i\alpha}A_{\alpha\beta}^{-1}B_{l\beta}B_{j\gamma}A_{\gamma\delta}^{-1}\partial_l B_{k\delta}\\
&& +c_{il}B_{j\alpha}B_{k\beta}\partial_l A_{\alpha\beta}^{-1}+\bar{J}_{il}B_{j\alpha}B_{k\beta}\partial_l A_{\alpha\beta}^{-1} +
B_{i\alpha}A_{\alpha\beta}^{-1}B_{l\beta}B_{j\gamma}B_{k\delta}\partial_l A_{\gamma\delta}^{-1}\\
&&+ B_{i\alpha}A_{\alpha\beta}^{-1}B_{l\beta}\partial_l \bar{J}_{jk}-B_{il}\partial_l \bar{J}_{jk}+\circlearrowleft_{(i,j,k)},
\end{eqnarray*}
where we have also replaced $B_{i\alpha}B_{j\gamma}\partial_l B_{k\delta}$ by $B_{k\alpha}B_{i\gamma}\partial_l B_{j\delta}$ by a circular permutation of $(i,j,k)$.
Inserting Eq.~(\ref{eqn:Jnn2}) inside $S_{ijk}$ gives
\begin{eqnarray}
S_{ijk}&=&(-B_{il}\partial_l B_{j\alpha}+B_{jl}\partial_l B_{i\alpha}+B_{l\alpha}\partial_l \bar{J}_{ij}-A_{\alpha l}\partial_l \bar{J}_{ij})A_{\alpha\beta}^{-1}B_{k\beta}\nonumber \\
&& +c_{il}B_{j\alpha}B_{k\beta}\partial_l A_{\alpha\beta}^{-1}+\bar{J}_{il}B_{j\alpha}B_{k\beta}\partial_l A_{\alpha\beta}^{-1} +
B_{i\alpha}A_{\alpha\beta}^{-1}B_{l\beta}B_{j\gamma}B_{k\delta}\partial_l A_{\gamma\delta}^{-1}\nonumber \\
&& +B_{i\alpha}A_{\alpha\beta}^{-1}B_{l\beta}\partial_l B_{j\gamma}A_{\gamma\delta}^{-1} B_{k\delta}+ B_{i\alpha}A_{\alpha\beta}^{-1}B_{l\beta}B_{j\gamma}A_{\gamma\delta}^{-1}\partial_l B_{k\delta} \nonumber \\
 && + B_{i\alpha}A_{\alpha\beta}^{-1}B_{l\beta}\partial_l \bar{J}_{jk}-B_{il}\partial_l \bar{J}_{jk}+\circlearrowleft_{(i,j,k)}. \label{eqn:Sijktemp}
\end{eqnarray}
Since $A_{\alpha l}\partial_l \bar{J}_{ij}A_{\alpha\beta}^{-1}B_{k\beta}=-B_{kl}\partial_l \bar{J}_{ij}$, the term $B_{kl}\partial_l \bar{J}_{ij}$ cancels with $-B_{il}\partial_l \bar{J}_{jk}$ by using a permutation on the indices $i,j,k$. Similarly, the term $B_{k\beta}A_{\alpha\beta}^{-1}B_{l\alpha}\partial_l \bar{J}_{ij}$ cancels with $B_{i\alpha}A_{\alpha\beta}^{-1}B_{l\beta}\partial_l \bar{J}_{jk}$ with a permutation of the indices $i,j,k$ and the antisymmetry of ${\mathbb A}^{-1}$. Next, we rewrite the terms 
$$
U_{ijk}\equiv B_{i\alpha}A_{\alpha\beta}^{-1}B_{l\beta}\partial_l B_{j\gamma}A_{\gamma\delta}^{-1} B_{k\delta}+ B_{i\alpha}A_{\alpha\beta}^{-1}B_{l\beta}B_{j\gamma}A_{\gamma\delta}^{-1}\partial_l B_{k\delta}+\circlearrowleft_{(i,j,k)},
$$
into
$$
U_{ijk}=B_{i\alpha}A_{\alpha\beta}^{-1}A_{\gamma\delta}^{-1} B_{k\delta}\left(
B_{l\beta}\partial_l B_{j\gamma}-B_{l\gamma}\partial_l B_{j\beta}\right) +\circlearrowleft_{(i,j,k)},
$$
using a circular permutation of $(i,j,k)$ and a relabeling of the silent indices $(\alpha,\beta,\gamma,\delta)$ as well as the antisymmetry of ${\mathbb A}^{-1}$. 
Inserting Eq.~(\ref{eqn:Jn22}) gives
\begin{eqnarray*}
U_{ijk} &=& B_{il}A_{\gamma\delta}^{-1}B_{k\delta}\partial_l B_{j\gamma}+B_{i\alpha}A_{\alpha\beta}^{-1}B_{kl}\partial_l B_{j\beta}\\
&& + B_{i\alpha}A_{\alpha\beta}^{-1}A_{\gamma\delta}^{-1}B_{k\delta}c_{jl}\partial_l A_{\gamma \beta}+B_{i\alpha}A_{\alpha\beta}^{-1}A_{\gamma\delta}^{-1}B_{k\delta}B_{jl}\partial_l A_{\gamma \beta}\\
&& +B_{i\alpha}A_{\alpha\beta}^{-1}A_{\gamma\delta}^{-1}B_{k\delta}\bar{J}_{jl}\partial_l A_{\gamma\beta}+\circlearrowleft_{(i,j,k)}.
\end{eqnarray*}
From the identity $A_{\beta\gamma}^{-1}\partial_l A_{\alpha\beta}=-A_{\alpha\beta}\partial_l A_{\beta\gamma}^{-1}$ (which comes from differentiating ${\mathbb A}{\mathbb A}^{-1}={\mathbb 1}_{k_2}$), we notice the following cancellations using circular permutations of $(i,j,k)$: The term $B_{i\alpha}A_{\alpha\beta}^{-1}A_{\gamma\delta}^{-1}B_{k\delta}c_{jl}\partial_l A_{\gamma\beta}$ cancels with $c_{il}B_{j\alpha}B_{k\beta}\partial_l A_{\alpha\beta}^{-1}$, and the same holds for $B_{i\alpha}A_{\alpha\beta}^{-1}A_{\gamma\delta}^{-1}B_{k\delta}\bar{J}_{jl}\partial_l A_{\gamma\beta}$ which cancels with $\bar{J}_{il}B_{j\alpha}B_{k\beta}\partial_l A_{\alpha\beta}^{-1}$. The first two terms in $U_{ijk}$ cancel with the first two terms of $S_{ijk}$ in Eq.~(\ref{eqn:Sijktemp}). It follows that 
\begin{equation}
\label{eqn:Sijkfin}
S_{ijk}=B_{i\alpha}B_{j\gamma}B_{k\delta}B_{l\beta} A_{\alpha\beta}^{-1}\partial_l A_{\gamma\delta}^{-1}+B_{i\alpha}B_{k\delta}A_{\alpha\beta}^{-1}A_{\gamma\delta}^{-1}B_{jl} \partial_l A_{\gamma\beta}+\circlearrowleft_{(i,j,k)}.
\end{equation}

By differentiating ${\mathbb A}{\mathbb A}^{-1}={\mathbb 1}_{k_2}$ and using the antisymmetry of ${\mathbb A}$, we have
$$
\partial_l A_{\gamma\delta}^{-1}=A_{\gamma m}^{-1}A_{\delta n}^{-1}\partial_l A_{mn}.
$$
Inserting this expression into the first term of $S_{ijk}$ in Eq.~(\ref{eqn:Sijkfin}) gives 
\begin{eqnarray*}
S_{ijk}&=&B_{i\alpha}B_{j\gamma}B_{k\delta} A_{\alpha\beta}^{-1}A_{\delta m}^{-1}A_{n\gamma}^{-1} \left( B_{l\beta} \partial_l A_{mn}+\circlearrowleft_{(\beta,n,m)}\right)\\
&&  + \left( B_{i\alpha}B_{k\delta}A_{\alpha\beta}^{-1}A_{\gamma\delta}^{-1}B_{jl} \partial_l A_{\gamma\beta}+\circlearrowleft_{(i,j,k)}\right),
\end{eqnarray*}
since
$$
B_{i\alpha}B_{j\gamma}B_{k\delta}B_{l\beta}A_{\alpha\beta}^{-1}\partial_l A_{\gamma\delta}^{-1}+\circlearrowleft_{(i,j,k)}=B_{i\alpha}B_{j\gamma}B_{k\delta}
\left( B_{l\beta}A_{\alpha\beta}^{-1}\partial_l A_{\gamma\delta}^{-1}+\circlearrowleft_{(\alpha,\gamma,\delta)}\right).
$$
Now we use Eq.~(\ref{eqn:J222}) from which we obtain
\begin{eqnarray*}
S_{ijk}&=&B_{i\alpha}B_{j\gamma}B_{k\delta} \left(A_{\gamma m}^{-1}A_{\delta n}^{-1} \partial_\alpha A_{mn}+\circlearrowleft_{(\alpha,\gamma,\delta)}\right)\\
&&  + \left( B_{i\alpha}B_{k\delta}A_{\alpha\beta}^{-1}A_{\gamma\delta}^{-1}B_{jl} \partial_l A_{\gamma\beta}+\circlearrowleft_{(i,j,k)}\right),\\
&=& B_{i l}B_{j\gamma}B_{k\delta} A_{\gamma m}^{-1}A_{\delta n}^{-1} \partial_l A_{mn}+ B_{i\alpha}B_{k\delta}A_{\alpha\beta}^{-1}A_{\gamma\delta}^{-1}B_{jl} \partial_l A_{\gamma\beta}+\circlearrowleft_{(i,j,k)}.
\end{eqnarray*}
By using a circular permutation of $(i,j,k)$, a relabeling of the silent indices and the antisymmetry of ${\mathbb A}^{-1}$, we show that the two terms in the last equation for $S_{ijk}$ cancel each other. Therefore $S_{ijk}=0$, and the Jacobi identity is satisfied. 

\end{itemize} 

\section{Non-unicity of Dirac brackets} 

Conditions~(\ref{cond1})-(\ref{cond3}) on the determination of the pseudoinverse ${\mathbb D}$ corresponds to a unique ${\mathbb D}$ for finite dimensional Hamiltonian systems. A natural question is whether or not one of these conditions can be relaxed with the requirement that the resulting matrix ${\mathbb J}_*$ given by Eq.~(\ref{eqn:Jstarmat}) still satisfies the Jacobi identity. In other words, what is the minimal set of equations which has to be satisfied by ${\mathbb D}$ such that the resulting bracket is a Poisson bracket and the second class constraints are Casimir invariants? 

We consider a general (antisymmetric) matrix ${\mathbb D}$ written in the same coordinates as in the previous section~:
$$
{\mathbb D}=\left(\begin{array}{ccc} 
D^{(11)} & D^{(12)} & D^{(13)}\\ -D^{(12)\dagger} & D^{(22)} & D^{(23)}\\
-D^{(13)\dagger} & -D^{(23)\dagger} & D^{(33)}
\end{array}\right).
$$
From the computation of ${\mathbb J}_*$ given by Eq.~(\ref{eqn:Jstar}), we notice that the resulting Dirac bracket is independent of $D^{(12)}$, $D^{(22)}$ and $D^{(23)}$. The second line and second column correspond to the Casimir invariants of the original Poisson bracket. In what follows, we choose these matrices to be zero. In addition, a natural choice for $D^{(33)}$ is ${\mathbb A}^{-1}$ in order to resolve the second class constraints. As a result, the matrix ${\mathbb J}_*$ is given by
$$
{\mathbb J}^*=\left( \begin{array}{cccc} 0 & 0 & 0 & -c^\dagger\\
0 & 0 & 0 & 0 \\ 0 & 0 & 0 & -{\mathbb A}D^{(13)\dagger} c^\dagger\\ 
c & 0 & -cD^{(13)}{\mathbb A} & \bar{\mathbb J}_*  \end{array}\right),
$$
where $\bar{\mathbb J}_*=\bar{\mathbb J}+{\mathbb B}{\mathbb A}^{-1}{\mathbb B}^\dagger +cD^{(11)}c^\dagger+cD^{(13)}{\mathbb B}^\dagger-{\mathbb B}D^{(13)\dagger} c^\dagger$. This matrix ${\mathbb J}_*$ does not satisfy the Jacobi identity in general. The goal is to find the conditions on $D^{(11)}$ and $D^{(13)}$ such that the resulting Dirac bracket satisfies the Jacobi identity regardless of the specific form of the original Poisson bracket (i.e., for all $c$, ${\mathbb A}$, ${\mathbb B}$ and $\bar{\mathbb J}$ satisfying Eqs.~(\ref{eqn:J11n})-(\ref{eqn:Jnnn})). For this purpose we first consider $S_{ijk}$ given by Eq.~(\ref{Jacobiator}) for $i\in I_2$ and $j,k\in I_{\mathrm n}$ and look at the terms proportional to $D^{(13)}_{\alpha\beta}$. After some algebra, the condition $S_{ijk}=0$ reduces to 
$$
D^{(13)}_{\alpha\beta}\left[c_{k\alpha}c_{jl}-c_{j\alpha}c_{kl}+c_{k\alpha} (\bar{\mathbb J}+{\mathbb B}{\mathbb A}^{-1}{\mathbb B}^\dagger)_{jl}-c_{j\alpha}(\bar{\mathbb J}+{\mathbb B}{\mathbb A}^{-1}{\mathbb B}^\dagger)_{kl}\right]\partial_l A_{i\beta}=0. 
$$ 
Since the coefficient of $D^{(13)}_{\alpha\beta}$ is non-zero in general, it implies that $D^{(13)}=0$. The condition $D^{(13)}=0$ implies that the second class constraints are Casimir invariants of the Dirac bracket~(\ref{eqn:gDB}). In other terms it is necessary to have the second class constraints as Casimir invariants in order to satisfy the Jacobi identity for the bracket~(\ref{eqn:gDB}). This result echoes the one in Ref.~\cite{chan13} where it was proven that if all the constraints are Casimir invariants (which is the case for a complete reduction) then the Dirac bracket is a Poisson bracket, i.e., it satisfies the Jacobi identity everywhere in phase space.  

Concerning $D^{(11)}$, there are two conditions to be satisfied~:
\begin{itemize}
\item The first condition is given by $S_{ijk}=0$ when $i\in I_1$ and $j,k\in I_{\mathrm n}$. This condition reduces to 
$$
S_{ijk}=-c_{li}\partial_l (cD^{(11)}c^\dagger)_{jk}-(cD^{(11)}c^\dagger)_{kl}\partial_l c_{ji}
+(cD^{(11)}c^\dagger)_{jl}\partial_l c_{ki}=0. 
$$
The terms proportional to $D^{(11)}_{\alpha\beta}$ vanish due to Eq.~(\ref{eqn:J11n}) and the antisymmetry of $D^{(11)}$. What is left are terms proportional to the derivatives of $D^{(11)}$~:
$$
S_{ijk}=-c_{j\alpha}c_k\beta c_{l i}\partial_l D^{(11)}_{\alpha\beta}. 
$$ 
Since $S_{ijk}$ should vanish regardless of $c$, it implies that $D^{(11)}$ does not depend on the variables $z_l$ for $l\in I_{\mathrm n}$ (i.e., the variables which are neither constraints nor Casimir invariants).  
\item The second condition is obtained for $i,j,k\in I_{\mathrm n}$ as
\begin{eqnarray*}
S_{ijk}&=&c_{il}\partial_l (cD^{(11)}c^\dagger)_{jk}+(cD^{(11)}c^\dagger)_{il}\partial_l (\bar{\mathbb J}+{\mathbb B}{\mathbb A}^{-1}{\mathbb B}^\dagger)_{jk}\\
&& +(\bar{\mathbb J}+{\mathbb B}{\mathbb A}^{-1}{\mathbb B}^\dagger)_{il}\partial_l (cD^{(11)}c^\dagger)_{jk}+(cD^{(11)}c^\dagger)_{il}\partial_l (cD^{(11)}c^\dagger)_{jk}+\circlearrowleft_{(i,j,k)}. 
\end{eqnarray*}
This decomposes in three series of terms, the ones linearly proportional to the coefficients of $D^{(11)}$, the ones proportional to the derivatives of these coefficients, and the ones which are quadratic in the coefficients of $D^{(11)}$ and their derivatives. The first series of terms is equal to 
$$
S_{ijk}^{(1)}= D^{(11)}_{\alpha\beta} \left[ c_{il}\partial_l (c_{j\alpha}c_{k\beta}) +c_{i\alpha}c_{l\beta}\partial_l (\bar{\mathbb J}+{\mathbb B}{\mathbb A}^{-1}{\mathbb B}^\dagger)_{jk}+ (\bar{\mathbb J}+{\mathbb B}{\mathbb A}^{-1}{\mathbb B}^\dagger)_{il}\partial_l (c_{j\alpha}c_{k\beta})\right]+\circlearrowleft_{(i,j,k)}. 
$$
Using Eq.~(\ref{eqn:intermDB}) together with a circular permutation of the indices $(i,j,k)$ and the antisymmetry of $D^{(11)}$, all these terms vanish, i.e., $S_{ijk}^{(1)}=0$. For the second series of terms, we have
$$
S_{ijk}^{(2)}=c_{j\alpha}c_{k\beta}\left(c_{il}+ (\bar{\mathbb J}+{\mathbb B}{\mathbb A}^{-1}{\mathbb B}^\dagger)_{il}\right) \partial_l D^{(11)}_{\alpha\beta}+\circlearrowleft_{(i,j,k)}.
$$ 
These terms are in general non-zero, except when $D^{(33)}$ does not depend on the variables $z_l$ for $l\in I_1$ and $l\in I_{\mathrm n}$. As for the third series of terms, it is written as
$$
S_{ijk}^{(3)}=c_{i\alpha}c_{j\gamma}c_{k\delta}c_{l\beta} D^{(11)}_{\alpha\beta}\partial_l D^{(11)}_{\gamma\delta}+\circlearrowleft_{(i,j,k)}.
$$
The conditions $S_{ijk}^{(3)}=0$ implies that $D^{(11)}$ does not depend on $z_l$ for $l\in I_{\mathrm n}$, which is again the condition found in the previous case $i\in I_1$ and $j,k\in I_{\rm n}$. 
\end{itemize}

In summary, the Jacobi identity is satisfied when $D^{(13)}=0$ and when $D^{(11)}$ does not depend on the variables $z_l$ for $l\in I_1$ and $l\in I_{\mathrm n}$, i.e., $D^{(11)}$ could have a dependence on the Casimir invariants of the bracket $\{\cdot,\cdot\}_*$ (whether these invariants originate from the Casimir invariants of the original bracket $\{\cdot,\cdot\}$ or from the reduction of the second-class constraints). Of course, depending on the specific choice of Poisson matrix ${\mathbb J}$, less restrictive cases can be considered for $D^{(13)}$ or $D^{(11)}$. An example is provided in Sec.~\ref{sec:example}. Here we provided the conditions on these two matrices such that the Jacobi identity is satisfied regardless of the specific form of the original Poisson matrix. 

Consequently, given that there are other possible choices for $D^{(11)}$, the Dirac bracket~(\ref{eqn:gDB}) as a Poisson bracket is not unique even if all the second class constraints are imposed as Casimir invariants. We notice that with a non-zero $D^{(11)}$, the resulting matrix ${\mathbb D}$ satisfies the conditions~(\ref{cond1}) and (\ref{cond3}), but the condition~(\ref{cond2}) is not satisfied. As a consequence, ${\cal P}_*=1-\hat{\cal Q}^\dagger {\mathbb D}\hat{\cal Q} {\mathbb J}$ is not a projector (with the additional assumption that $c^\dagger$ does not belong to the kernel of $D^{(11)}$). 

\section{Reduced dynamics}

We consider the Dirac-Poisson bracket given by the matrix
$$
{\mathbb J}^*=\left( \begin{array}{cccc} 0 & 0 & 0 & -c^\dagger\\
0 & 0 & 0 & 0 \\ 0 & 0 & 0 & 0\\ 
c & 0 & 0 & \bar{\mathbb J}_*  \end{array}\right),
$$
where $\bar{\mathbb J}_*=\bar{\mathbb J}+{\mathbb B}{\mathbb A}^{-1}{\mathbb B}^\dagger +cD^{(11)}c^\dagger$. The equations of motion for $z_l$ for $l\in I_1$ and $l\in I_{\rm n}$ are given by
\begin{eqnarray*}
&& \dot{z_1}=-c^\dagger \frac{\partial H}{\partial z_{\rm n}} ,\\
&& \dot{z_{\rm n}} = \bar{\mathbb J}_* \frac{\partial H}{\partial z_{\rm n}} +c \frac{\partial H}{\partial z_{\rm 1}},
\end{eqnarray*}
where $H$ is the Hamiltonian which a priori depends on $z_1$ and $z_{\rm n}$ (since $z_2$ and $z_{\rm c}$ are Casimir invariants, they can be forgotten in the analysis). 
In order to have a complete reduction, the first-class constraints have to be conserved quantities (even if they are not Casimir invariants) which is obtained under the condition
\begin{equation}
\label{eq:redcond}
c^\dagger  \frac{\partial H}{\partial z_{\rm n}}=0.
\end{equation}
Under this condition, we notice that the resulting system of dynamical equations is unique, i.e., it does not depend on $D^{(11)}$ since the additional term in the equations $cD^{(11)}c^\dagger \partial H/\partial z_{\rm n}$ vanishes.
 
The condition~(\ref{eq:redcond}) is rewritten as $\{z_1,H\}_*=0$ since $\{z_1,z_2\}_*=\{z_1,z_{\rm c}\}_*=0$ (where $z_2$ denotes $z_l$ with $l\in I_2$ and $z_{\rm c}$ denotes $z_l$ with $l\in I_{\rm c})$, which means that the Hamiltonian after the reduction has to commute with the first-class constraints. In other terms, the Hamiltonian has to be changed so that the first-class constraints are conserved quantities of the Dirac bracket. 

\section{Examples}
\label{sec:example}

There are two trivial examples and they correspond to the two limits of the Dirac reduction, one for which all the constraints are first-class and another for which all the constraints are second-class. 
For the case where all the constraints are first-class, the matrix ${\mathbb C}$ is zero and its pseudoinverse ${\mathbb D}$ is also zero. Hence the Dirac bracket is the same as the original bracket, which means that the reduction has failed. 
The second limit example corresponds to the case where the matrix ${\mathbb C}$ is invertible. The reduction is complete and the Dirac bracket is the usual one. 

Between these two limit cases, there are examples where the reduction is incomplete. Several examples are given below.  

\subsection{Finite dimensional examples}

If we consider only one constraint, it is necessary a first-class constraint since it Poisson commutes with itself. Therefore the Dirac bracket computed with the pseudoinverse is identical to the original Poisson bracket. 

If we consider two constraints, there are two cases: one in which they do not commute and hence both constraints are second class, and another one where they commute and hence they are both first-class. In the first case, the Dirac bracket is identical to the usual one since the matrix ${\mathbb C}$ is invertible. In the second case, the Dirac bracket is identical to the original one since ${\mathbb C}=0$. 

More interesting cases occur when there are three constraints, one of which is first-class. The simplest example is afforded by the Poisson bracket
$$
{\mathbb J}=\left(\begin{array}{cccc} 0 & 0 & 0 & 1\\
0 & 0 & -1 & 1\\ 0 & 1 & 0 & -1 \\ -1 & -1 & 1 & 0
\end{array} \right),
$$  
with the three constraints $\Phi_1({\bf z})=z_1$, $\Phi_2({\bf z})=z_2$ and $\Phi_3({\bf z})=z_3$. We notice that $\Phi_1$ is a first-class constraint whereas $\Phi_2$ and $\Phi_3$ are second-class. The matrix ${\mathbb C}$ is given by
$$
{\mathbb C}=\left(\begin{array}{cccc} 0 & 0 & 0 \\
0 & 0 & -1\\ 0 & 1 & 0 
\end{array} \right),
$$ 
and its pseudoinverse by
$$
{\mathbb D}=\left(\begin{array}{cccc} 0 & 0 & 0 \\
0 & 0 & 1\\ 0 & -1 & 0 
\end{array} \right).
$$
The Dirac bracket computed from the pseudoinverse is given by
$$
{\mathbb J}_*=\left(\begin{array}{cccc} 0 & 0 & 0 & 1\\
0 & 0 & 0 & 0\\ 0 & 0 & 0 & 0 \\ -1 & 0 & 0 & 0
\end{array} \right),
$$ 
where we notice that $\Phi_2$ and $\Phi_3$ are Casimir invariants of ${\mathbb J}_*$, but $\Phi_1$ is not. 

The second example in finite dimensions is given by 
$$
{\mathbb J}=\left(\begin{array}{cccccc} 0 & -z_3 & z_2 & 0 & -w_3 & w_2\\
z_3 & 0 & -z_1 & w_3 & 0 & -w_1\\ -z_2 & z_1 & 0 & -w_2 & w_1 & 0 \\ 0 & -w_3 & w_2 & 0 & 0 & 0\\ w_3 & 0 & -w_1 & 0 & 0 & 0\\ -w_2 & w_1 & 0 & 0 & 0 & 0
\end{array} \right),
$$  
which corresponds to the Poisson matrix for the rigid body~\cite{Bmars02}, and three constraints given by $\Phi_1({\bf z},{\bf w})=z_3$, $\Phi_2({\bf z},{\bf w})=w_2$ and $\Phi_3({\bf z},{\bf w})=w_3$. We notice that $\Phi_3$ is a first-class constraint. The matrix ${\mathbb C}$ is given by
$$
{\mathbb C}=\left(\begin{array}{cccc} 0 & w_1 & 0 \\
-w_1 & 0 & 0\\ 0 & 0 & 0 
\end{array} \right),
$$ 
and its pseudoinverse by
$$
{\mathbb D}=\left(\begin{array}{cccc} 0 & -1/w_1 & 0 \\
1/w_1 & 0 & 0\\ 0 & 0 & 0 
\end{array} \right).
$$
The Dirac bracket computed from the pseudoinverse is given by
\begin{equation}
\label{eq:Jstarex2}
{\mathbb J}_*=\left(\begin{array}{cccccc} 0 & z_1w_3/w_1-z_3 & 0 & -w_2w_3/w_1 & 0 & w_2\\
-z_1w_3/w_1+z_3 & 0 & 0 & w_3 & 0 & -w_1\\ 0 & 0 & 0 & 0 & 0 & 0 \\ w_2w_3/w_1 & -w_3 & 0 & 0 & 0 & 0 \\ 0 & 0 & 0 & 0 & 0 & 0\\ -w_2 & w_1 & 0 & 0 & 0 & 0
\end{array} \right),
\end{equation}
with the same remark as above, $\Phi_3$ is not a Casimir invariant of the Dirac bracket. 

In order to illustrate the remark above on the unicity of the Dirac bracket, we consider matrices ${\mathbb D}$ which only satisfy condition~(\ref{cond1}). For instance a possible matrix is 
$$
{\mathbb D}=\left(\begin{array}{cccc} 0 & -1/w_1 & 0 \\
1/w_1 & 0 & -w_2\\ 0 & w_2 & 0 
\end{array} \right),
$$
which satisfies ${\mathbb C}={\mathbb C}{\mathbb D}{\mathbb C}$ and ${\mathbb D}={\mathbb D}{\mathbb C}{\mathbb D}$ but not ${\mathbb C}{\mathbb D}={\mathbb D}{\mathbb C}$. 
This choice leads to the Dirac bracket with the following Poisson matrix~:
$$
{\mathbb J}_*=\left(\begin{array}{cccccc} 0 & -w_1w_2w_3+z_1w_3/w_1-z_3 & w_1w_2^2 & -w_2w_3/w_1 & 0 & w_2\\
w_1w_2w_3-z_1w_3/w_1+z_3 & 0 & -w_1^2w_2 & w_3 & 0 & -w_1\\ -w_1w_2^2 & w_1^2w_2 & 0 & 0 & 0 & 0 \\ w_2w_3/w_1 & -w_3 & 0 & 0 & 0 & 0 \\ 0 & 0 & 0 & 0 & 0 & 0\\ -w_2 & w_1 & 0 & 0 & 0 & 0
\end{array} \right),
$$
which is different from ${\mathbb J}_*$ given by Eq.~(\ref{eq:Jstarex2}). 

\subsection{Infinite dimensional example: two-dimensional Euler's equation}
 
For an infinite-dimensional Hamiltonian system with dynamical field variables ${\bm\chi}({\bf x})$, the generalized Dirac bracket is written as
$$
\{F,G\}_*=\{F,G\}-\int d^N x \int d^N x' \{F,\Phi_n({\bf x})\} D_{nm}({\bf x},{\bf x}') \{\Phi_m({\bf x}'),G\},
$$ 
where ${\mathbb D}$ satisfies
\begin{eqnarray*}
&& C_{nm}({\bf x},{\bf x}')=\int d^N y \int d^N y' C_{nl}({\bf x},{\bf y})D_{lp}({\bf y},{\bf y}')C_{pm}({\bf y}',{\bf x}'),\\
&& D_{nm}({\bf x},{\bf x}')=\int d^N y \int d^N y' D_{nl}({\bf x},{\bf y})C_{lp}({\bf y},{\bf y}')D_{pm}({\bf y}',{\bf x}'),\\
&& \int d^N y C_{nl}({\bf x},{\bf y})D_{lm}({\bf y},{\bf x}')=\int d^N y D_{nl}({\bf x},{\bf y})C_{lm}({\bf y},{\bf x}'),
\end{eqnarray*}
which generalize the conditions~(\ref{cond1})-(\ref{cond3}). The existence of such a pseudoinverse ${\mathbb D}$ is more complicated than in the finite dimensional case, and is beyond the scope of the present work. On a practical basis it has to be assessed case by case. In the infinite dimensional case, it is very difficult to separate first class and second class constraints as we shall see it below using an example. However it should be noted that the principles of the definition of the Dirac bracket does not rely on the separation between first class and second class constraints.  
 
Here we consider the two dimensional Euler equations for the density $\rho({\bf x})$, the fluid velocity ${\bf v}({\bf x})$ and the entropy $s({\bf x})$ where ${\bf x}=(x,y)\in {\mathbb T}^2$ (the two-dimensional torus) given by
\begin{eqnarray*}
&& \dot{\rho}=-\nabla\cdot (\rho {\bf v}),\\
&& \dot{\bf v}=-{\bf v}\cdot \nabla {\bf v}-\rho^{-1}\nabla P,\\
&& \dot{s}=-{\bf v}\cdot \nabla s.
\end{eqnarray*}
The Hamiltonian of the system is a function of the dynamical field variables $\rho$, ${\bf v}$ and $s$, given by
\begin{equation}
\label{eq:Hrhos}
H[\rho,{\bf v},s]=\int d^2x \left(\rho \frac{v^2}{2}+\rho U(\rho,s)\right),
\end{equation}
where $U$ is the internal energy such that the pressure is given by $P=\rho^2\partial U/\partial \rho$, and the Poisson bracket is
\begin{eqnarray}
\{F,G\}&=&-\int d^2x \left[ F_\rho \nabla \cdot G_{\bf v}+F_{\bf v}\cdot \nabla G_\rho -\rho^{-1} (\nabla\times {\bf v}) \cdot F_{\bf v}\times G_{\bf v}\right. \nonumber \\
&&\left. \qquad \qquad +\rho^{-1}\nabla s \cdot (F_s G_{\bf v}-F_{\bf v} G_s)\right],
\label{eqn:PBEuler}
\end{eqnarray}
where $F_\rho$ denotes the functional derivative of the observable $F$ with respect to the field variable $\rho$ (and equivalently for $F_{\bf v}$ and $F_s$).   
The associated Poisson operator is
$$
{\mathbb J}=\left( \begin{array}{ccc}
0 & -\nabla \cdot & 0\\
-\nabla & -\rho^{-1}(\nabla\times {\bf v})\times & \rho^{-1} \nabla s\\
0 & -\rho^{-1}\nabla s & 0
\end{array}\right),
$$
from which the Poisson bracket is defined as 
$$
\{F,G\}=\int d^2x F_{\bm \chi}\cdot {\mathbb J}G_{\bm\chi},
$$
where $F_{\bm \chi}=(F_\rho, F_{\bf v},F_s)$. 

In order to find reduced systems defined from some constraints imposed on the above Hamiltonian system, a first obvious choice would be to impose $\rho({\bf x})=\rho_0({\bf x})$ with a prescribed density $\rho_0$. Since $\{\rho({\bf x}),\rho({\bf x}')\}=0$, these constraints are all first class constraints and not of them are Casimir invariants. The matrix ${\mathbb C}$ is zero and hence its pseudo-inverse is also zero. The associated Dirac bracket is the same as the original bracket. The reduction has failed. 

A second choice is to impose a local constraint as divergence-free velocity field which consists of an infinite number of constraints, $\Phi({\bf x})=\nabla\cdot {\bf v}({\bf x})=0$. The idea behind this calculation is to have a divergence free velocity field (incompressibility) but keeping the density as a dynamical field variable. It should be noted that none of these constraints are Casimir invariants of the bracket~(\ref{eqn:PBEuler}). 

The operator ${\mathbb C}=\hat{\cal Q}{\mathbb J}\hat{\cal Q}^\dagger$ where $\hat{\cal Q}=(0,\nabla \cdot , 0)$, is given by
$$
{\mathbb C}=\nabla\cdot (\rho^{-1}(\nabla\times{\bf v})\times \nabla).
$$
We rewrite this operator using $q(x,y)=\rho^{-1}(\nabla\times {\bf v})\cdot \hat{\bf z}$~:
$$
{\mathbb C}=-[q,\cdot],
$$
where the bracket $[\cdot,\cdot]$ is given by
$$
[f,g]=\frac{\partial f}{\partial x}\frac{\partial g}{\partial y}-\frac{\partial f}{\partial y}\frac{\partial g}{\partial x}.
$$
The operator ${\mathbb C}$ is not invertible since any function of $q$ belongs to its kernel. Therefore the constraints form a complicated admixture of first-class and second-class constraints. In order to see whether or not ${\mathbb C}$ has a pseudoinverse, we consider the characteristics associated with the linear operator ${\mathbb C}$ which are fictitious trajectories with $q(x,y)$ as Hamiltonian. In this fictitious dynamics, the variables are $x$ and $y$. Locally, we change variables to action-angle variables $(\varphi, A)$ so that $q(x,y)=q_0(A)$. Since this change of fictitious variables is canonical, the expression of the linear operator ${\mathbb C}$ becomes
$$
{\mathbb C}=q'_0(A)\frac{\partial}{\partial\varphi}.
$$   
In Fourier series, it is straightforward to see that ${\mathbb C}$ has a pseudoinverse ${\mathbb D}$, and the action of these two operators are given by
\begin{eqnarray*}
&& {\mathbb C}f=q'_0(A)\sum_k ik f_k {\rm e}^{ik\varphi},\\
&& {\mathbb D}f=\frac{1}{q'_0(A)}\sum_{k\not= 0} \frac{f_k}{ik} {\rm e}^{ik \varphi}.
\end{eqnarray*}  
The Poisson operator ${\mathbb J}_*$ associated with the Dirac bracket defined from the pseudoinverse ${\mathbb D}$ is given by
$$
{\mathbb J}_*=\left(\begin{array}{ccc}
\Delta {\mathbb D}\Delta & -\nabla \cdot {\mathbb P} & -\Delta {\mathbb D}\nabla \cdot (\rho^{-1}\nabla s \bullet)\\
-{\mathbb P}^\dagger \nabla & -\rho^{-1}(\nabla\times {\bf v})\times {\mathbb P} & {\mathbb P}^\dagger (\rho^{-1} \nabla s \bullet)\\
\rho^{-1}\nabla s \cdot \nabla {\mathbb D}\Delta & -\rho^{-1}\nabla s\cdot {\mathbb P} & -\rho^{-1}\nabla s \cdot \nabla{\mathbb D}\nabla\cdot (\rho^{-1}\nabla s \bullet)
\end{array} \right),
$$
where $\bullet$ indicates where the operator acts, and ${\mathbb P}=1-\nabla {\mathbb D}\nabla \cdot (\rho^{-1}(\nabla\times {\bf v})\times \bullet)$ is a projector with the following properties~:
\begin{eqnarray*}
&& \nabla \times {\mathbb P}={\mathbb P},\\
&& {\mathbb P}\nabla = \nabla (1-{\mathbb D}{\mathbb C}),
\end{eqnarray*}
where $1-{\mathbb D}{\mathbb C}$ is the projector onto the kernel of ${\mathbb C}$. In fictitious action-angle variables, this projector is given by
$$
1-{\mathbb D}{\mathbb C}=\int d\varphi.
$$ 
From the Poisson matrix ${\mathbb J}_*$ the Dirac bracket becomes~:
\begin{eqnarray}
\label{eqn:DBEuler}
\{F,G\}_*&=&\int d^2x \left[ ( \nabla F_\rho -\rho^{-1}\nabla s F_s) \cdot {\bar{G_{\bf v}}}-{\bar{F_{\bf v}}}\cdot (\nabla G_\rho-\rho^{-1}\nabla s G_s)\right. \nonumber\\
&& \left. \qquad \qquad  +\rho^{-1} (\nabla\times {\bf v}) \cdot {\bar{F_{\bf v}}}\times  {\bar{G_{\bf v}}}\right],
\end{eqnarray}
where the constrained derivative $ {\bar{F_{\bf v}}}$ is defined from the Dirac projector ${\cal P}_*= 1-\hat{\cal Q}^\dagger{\mathbb D} \hat{\cal Q}{\mathbb J}$ as
$$
 {\bar{F_{\bf v}}}={\mathbb P} F_{\bf v}-\nabla {\mathbb D}\nabla\cdot (\nabla F_\rho-\rho^{-1}\nabla s F_s).
$$
For Hamiltonian~(\ref{eq:Hrhos}), the constrained derivative ${\bar{H_{\bf v}}}$ leads to the definition of a constrained velocity field $\bar{\bf v}$ such that ${\bar{H_{\bf v}}}=\rho\bar{\bf v}$ with 
$$
\bar{\bf v}={\bf v}-\rho^{-1}\nabla{\mathbb D}\nabla\cdot [({\bf v}\cdot\nabla){\bf v}+\rho^{-1}\nabla P].
$$ 
It should be noted that a priori $\nabla \cdot \bar{\bf v}$ is non-zero. 
Functionals $C$ satisfying the following conditions~:
\begin{eqnarray*}
&& \nabla C_\rho-\rho^{-1}\nabla s C_s=0,\\
&& {\mathbb P} C_{\bf v}=0,
\end{eqnarray*}
are Casimir invariants of the bracket~(\ref{eqn:DBEuler}).
The first condition leads to $C=\int d^2x \rho \psi(s)$ which are also Casimir invariants of the original bracket~(\ref{eqn:PBEuler}). The reduction procedure has preserved the Casimir invariants as expected. The second equation states that the second-class constraints are Casimir invariants. However due to the intricacy of first and second class constraints, it is cumbersome to explicit those invariants.  

The resulting equations of motion are the following ones~:
\begin{eqnarray*}
&& \dot{\rho}=-\nabla\cdot (\rho \bar{\bf v}),\\
&& \dot{\bf v}=-{\mathbb P}^\dagger \left( ({\bf v}\cdot\nabla) {\bf v}+\frac{1}{\rho}\nabla P\right),\\
&& \dot{s}=-\bar{\bf v}\cdot \nabla s.
\end{eqnarray*}

Not all the constraints have been fulfilled since
$$
\nabla\cdot \dot{\bf v}=-(1-{\mathbb C}{\mathbb D})\nabla \cdot \left[({\bf v}\cdot\nabla){\bf v}+\rho^{-1}\nabla P\right],
$$
which is in general non-zero since ${\mathbb C}$ is not invertible. In order to have a fully incompressible (in the sense of $\nabla\cdot {\bf v}=0$) model, the pressure needs to be adjusted such that $\nabla \cdot \dot{\bf v}=0$ when $\nabla \cdot {\bf v}=0$ as it is the case in the standard incompressible model. 
In summary, the Dirac procedure as outlined above provides a Dirac bracket which is a Poisson bracket, given by Eq.~(\ref{eqn:DBEuler}), but does not have all the constraints as Casimir invariants, contrary to the usual procedure where the matrix of the Poisson brackets between the various constraints is invertible. The origin of this incomplete reduction is the presence of first class constraints which are not Casimir invariants of the original bracket~\cite{chan13b}.

\section*{Acknowledgments}
This work was supported by the Agence Nationale de la Recherche (ANR GYPSI) and by the European Community under the contract of Association between EURATOM, CEA, and the French Research Federation for fusion study. The views and opinions expressed herein do not necessarily reflect those of the European Commission. CC acknowledges fruitful discussions with P.J. Morrison and with the \'Equipe de Dynamique Nonlin\'eaire of the Centre de Physique Th\'eorique of Marseille.


\section*{References}

\begin{thebibliography}{10}

\bibitem{dira50} P.A.M. Dirac, Can. J. Math. 2 (1950) 129. 

\bibitem{dira58} P.A.M. Dirac, Proc. Roy. Soc. Lond. A 246 (1958) 326.

\bibitem{Bhans76} A. Hanson, T. Regge, C. Teitelboim, Constrained Hamiltonian Systems, Accademia Nazionale dei Lincei, Roma, 1976.

\bibitem{Bsund82} K. Sundermeyer, Constrained Dynamics, Springer-Verlag, Berlin, 1982.

\bibitem{Bhenn92} M. Henneaux, C. Teitelboim, Quantization of Gauge Systems, Princeton University Press, Princeton, New Jersey, 1992. 

\bibitem{wipf94} A.W. Wipf, Hamilton's formalism for systems with constraints, in 
Canonical Gravity: From Classical to Quantum, Lecture Notes in Physics 434 (1994) 22.

\bibitem{morr09} P.J. Morrison, N. Lebovitz, J. Biello, Ann. Phys. 324 (2009) 1747.

\bibitem{chan10} C. Chandre, E. Tassi, P.J. Morrison, Phys. Plasmas 17 (2010) 042307.

\bibitem{chan12} C. Chandre, P.J. Morrison, E. Tassi, Phys. Lett. A 376 (2012) 737.

\bibitem{chan13} C. Chandre, L. de Guillebon, A. Back, E. Tassi, P.J. Morrison, J. Phys. A: Math. Theor. 46 (2013) 125203.

\bibitem{deri96} A.A. Deriglazov, A.V. Galajinsky, S.L. Lyakhovitch, Nucl. Phys. B 473 (1996) 245.

\bibitem{bizd00} C. Bizdadea, A. Constantin, S.O. Saliu, Europhys. Lett. 50 (2000) 169.

\bibitem{abre02} E.M.C. Abreu, D. Dalmazi, E.A. Silva, Int. J. Mod. Phys. A 17 (2002) 395. 

\bibitem{bizd07} C. Bizdadea, E.M. Cioroianu, S.O. Saliu, S.C. Sararu, O. Balus, J. Phys. A: Math. Theor. 40 (2007) 14537. 

\bibitem{Bmars02} J.E. Marsden, T.R. Ratiu, Introduction to Mechanics and Symmetry, Springer-
Verlag, Berlin, 2002.

\bibitem{morr82} P.J. Morrison, in Mathematical Methods in Hydrodynamics and Integrability in Related Dynamical Systems, La Jolla Institute, 1981, edited by M. Tabor and Y.M. Treve, AIP Conf. Proc. 88 (1982) 13.

\bibitem{morr98} P.J. Morrison, Rev. Mod. Phys. 70 (1998) 467.

\bibitem{morr80b} P.J. Morrison, Phys. Lett. A  80 (1980) 383.

\bibitem{mars82} J.E. Marsden, A. Weinstein, Physica D 4 (1982) 394.

\bibitem{dres88} A. Dresse, J. Fisch, M. Henneaux, C. Schomblond, Phys. Lett. B 210 (1988) 141.

\bibitem{chan13b} C. Chandre, J. Phys. A: Math. Theor. 46 (2013) 375201. 

\end{thebibliography}
\end{document}